\documentclass[11pt,onecolumn]{article}

\usepackage[dvips]{graphicx}
\DeclareGraphicsExtensions{.eps}
\usepackage{bm}
\usepackage{amsfonts}
\usepackage{amsmath}
\usepackage{amssymb}
\usepackage{cite}
\usepackage[amsmath,thmmarks]{ntheorem}

\usepackage{color, soul}
\usepackage{epic, eepic}
\usepackage[top=1in, bottom=1in, left=1.25in, right=1.25in]{geometry}
\usepackage{appendix}
\usepackage{color, soul}
\usepackage{array}
\usepackage{theorem}
\usepackage{booktabs}

\theoremheaderfont{\sc}\theorembodyfont{\upshape}
\theoremstyle{nonumberplain}
\theoremseparator{}
\theoremsymbol{\rule{1ex}{1ex}}
\newtheorem{proof}{Proof}

\begin{document}

%
\title{A stochastic gradient approach on compressive sensing signal reconstruction based on adaptive filtering framework}
%
%
%

\author{Jian~Jin,~
        Yuantao~Gu\thanks{This work was supported in part by the National Natural Science Foundation of China under Grants NSFC 60872087 and NSFC U0835003. The authors are with the Department of Electronic Engineering, Tsinghua University, Beijing 100084, China. The corresponding author of this paper is Yuantao Gu (e-mail: gyt@tsinghua.edu.cn).}
,~and~Shunliang~Mei~}

\date{Received Feb. 27, 2009; accepted Oct. 3, 2009.\\\vspace{1em}
This article appears in \textsl{IEEE Journal of Selected topics in Signal Processing}, 4(2):409-420, 2010.}

\maketitle

\begin{abstract}
Based on the methodological similarity between sparse signal
reconstruction and system identification, a new approach for sparse
signal reconstruction in compressive sensing (CS) is proposed in
this paper. This approach employs a stochastic gradient-based
adaptive filtering framework, which is commonly used in system
identification, to solve the sparse signal reconstruction problem.
Two typical algorithms for this problem: $l_0$-least mean square
($l_0$-LMS) algorithm  and $l_0$-exponentially forgetting window LMS
($l_0$-EFWLMS) algorithm are hence introduced here. Both the
algorithms utilize a zero attraction method, which has been
implemented by minimizing a continuous approximation of $l_0$ norm
of the studied signal. To improve the performances of these proposed
algorithms, an $l_0$-zero attraction projection ($l_0$-ZAP)
algorithm is also adopted, which has effectively accelerated their
convergence rates, making them much faster than the other existing
algorithms for this problem. Advantages of the proposed approach,
such as its robustness against noise etc., are demonstrated by
numerical experiments.
\end{abstract}

\section{Introduction}
\subsection{Overview of Compressive Sampling}
Compressive sensing or compressive sampling (CS)
\cite{CSbase,CS1,CS2,CS3} is a novel technique that enables sampling
below Nyquist rate, without (or with little) sacrificing
reconstruction quality. It is based on exploiting signal sparsity in
some typical domains. A brief review on CS is given here.

For a piece of finite-length, real-valued 1-D discrete signal
$\textbf{x}$, its representation in domain ~$\bf \Psi$~ is
\begin{equation}
{\bf x}=\sum_{i=1}^{N}{\bf{\psi}}_is_i={\bf \Psi s},
\end{equation}
where $\textbf{x}$ and $\textbf{s}$ are $N\times 1$ column
vectors, and $\bf \Psi$ is an $N\times N$ basis matrix with
vectors $\{{\bf \psi}_i\}(i=1,2,...,N)$ as columns. Obviously,
$\textbf{x}$ and $\textbf{s}$ are equivalent representations of
the signal when $\bf \Psi$ is full ranked. Signal $\textbf{x}$ is
$K$-sparse if $K$ out of $N$ coefficients of $\bf s$ are nonzero
 in the domain $\bf \Psi$. And it is sparse if $K\ll N$.

Take $M$ $(K\leq M \ll N)$ linear, non-adaptive measurement of
$\textbf{x}$ through a linear transform $\bf \Phi$, which is
\begin{equation}\label{Sampling}
    \bf y=\Phi x=\Phi \Psi s=As,
\end{equation}
where $\bf \Phi$ is an $M\times N$ matrix, and each of its
$M$ rows can be considered as a basis vector, usually orthogonal.
$\textbf{x}$ is thus transformed, or down sampled, to an $M\times
1$ vector $\textbf{y}$.

According to the discussion above, the main task of CS is
\begin{itemize}
  \item To design a stable measurement matrix. It is important to
  make a sensing matrix which allows recovery of as many
  entries of $\bf x$ as possible with as few as $M$ measurements.
  The matrix $\bf A$ should satisfy the conditions of Incoherence and restricted isometry property (RIP) \cite{CS2}.
  Fortunately, simple choice of $\bf \Phi$ as the random matrix can make $\bf A$ satisfy
  these conditions with high possibility. Common design methods include \emph{Gaussian
  measurements, Binary measurements, Fourier measurements}, and
  \emph{Incoherent measurement} \cite{CS2}. The Gaussian
  measurements are employed in this work, i.e., the entries of $M\times N$ sensing
  matrix $\bf \Phi$ are independently sampled from a normal distribution
  with mean zero and variance $1/M$ ($\mathcal{N}(0,1/M)$). When the basis matrix $\bf \Psi$
  (wavelet, Fourier, discrete cosine transform (DCT), etc) is orthogonal, $\bf A$ is also independent and identically-distributed
(i.i.d.) with $\mathcal{N}(0,1/M)$\cite{CS3}.
  \item To design a signal reconstruction algorithm. The signal reconstruction
  algorithm aims to find the sparsest solution to
  (\ref{Sampling}), which is ill-conditioned.
  This will be discussed in detail in the following subsection.
\end{itemize}

Compressive Sensing methods provide a robust framework that can
reduce the number of measurements required to estimate a sparse
signal. For this reason, CS methods are useful in many areas, such
as MR imaging \cite{MRI} and analog-to-digital conversion
\cite{AIC}.

\subsection{Signal Reconstruction Algorithms}
Although CS is a new concept emerged recently,
searching for the sparse solution to an under-determined system of
linear equations (\ref{Sampling}) has always been of significant
importance in signal processing and statistics. The main idea is
to obtain the sparse solution by adding sparse constraint. The
sparsest solution can be acquired by taking $l_0$ norm into
account,
\begin{equation}\label{l0constrint}
    \min_{\bf s} \|{\bf s}\|_0,~~~~ {\rm s.t.}~~~\bf{As=y}.
\end{equation}
Unfortunately, this criterion is not convex, and the computational
complexity of optimizing it is Non-Polynomial (NP) hard. To overcome
this difficulty, $l_0$ norm has to be replaced by simpler ones in
terms of computational complexity. For example, the convex $l_1$
norm is used,
\begin{equation}\label{l1constrint}
    \min_{\bf s} \|{\bf s}\|_1,~~~~ {\rm s.t.}~~~\bf{As=y}.
\end{equation}
This idea is known as {\textit{basis pursuit}}, and it can be
recasted as a linear programming (LP) issue. A recent body of
related research shows that perhaps there are conditions
guaranteeing a formal equivalence between the $l_0$ norm solution
and the $l_1$ norm solution \cite{CSbase}.

In the presence of noise and/or imperfect data, however, it is
undesirable to fit the linear system exactly. Instead, the
constraint in (\ref{l1constrint}) is relaxed to obtain the Basis
Pursuit De-Noise (BPDN) problem,
\begin{equation}\label{BPDN}
    \min_{\bf s} \|{\bf s}\|_1,~~~~ {\rm s.t.}~~~\|{\bf
    y-As}\|_2 \leq \sigma,
\end{equation}
where the positive parameter $\sigma$ is an estimation of the
noise level in the data. The convex optimization problem
(\ref{BPDN}) is one possible statement of the
least-squares problem regularized by
the $l_1$ norm. In fact, the BPDN label
is typically applied to the penalized least-squares problem,
\begin{equation}\label{QP}
    \min_{\bf s}~~~ \|{\bf y-As}\|_2^2+\lambda \|{\bf s}\|_1,
\end{equation}
which is proposed by Chen et al. in \cite{BPDN}, \cite{BPDN1}. The
third formulation,
\begin{equation}\label{Lasso}
    \min_{\bf s}~~~ \|{\bf y-As}\|_2^2~~~~ {\rm s.t.}~~~\|{\bf s}\|_1\leq
    \tau,
\end{equation}
which has an explicit $l_1$ norm constraint, is often called the
Least Absolute Shrinkage and Selection Operator (LASSO)
\cite{LASSO}. The problems (\ref{BPDN}), (\ref{QP}) and
(\ref{Lasso}) are identical in some situations. The
precise relationship among them is discussed in
\cite{Pursuitroot}, \cite{GP}.

Many approaches and their variants to these problems have been
described by the literature. They mainly fall into two basic
categories.

\textbf{Convex relaxation:} The first kind of convex optimization
methods to solve problems (\ref{BPDN}), (\ref{QP}) and
(\ref{Lasso}) includes interior-point (IP) methods \cite{IP1},
\cite{LNP}, which transfer these problems to a convex quadratic
problem. The standard IP methods cannot handle large scale
situation. However, many improved IP methods, which
exploit fast algorithms for the matrix vector operations with $\bf
A$ and ${\bf A}^{\rm T}$, can deal with large scale situation, as
demonstrated in \cite{BPDN}, \cite{IP}. High-quality
implementations of such IP methods include l1-magic \cite{l1magic}
and PDCO \cite{PDCO}, which use iterative algorithms, such as the
conjugate gradients (CG) or LSQR algorithm \cite{LSQR}, to compute
the search step. The fastest IP method has been recently proposed
to solve (\ref{QP}), different from the method used in the
previous works. In such method called $l1\_ls$, the search
operation in each step is done using the Preconditioned Conjugate
Gradient (PCG) algorithm, which requires less computation, i.e.,
only the products of $\bf A$ and ${\bf A}^{\rm T}$ \cite{l1ls}.

The second kind of convex optimization methods to solve problems
(\ref{BPDN}), (\ref{QP}) and (\ref{Lasso}) includes homotopy method
and its variants. Homotopy method is employed to find the full path
of solutions for all nonnegative values of the scalar parameters in
the above said three problems. When solution is extremely sparse,
the methods described in \cite{Osborne,Turlach,LARS} can be very
fast \cite{Fast}. Otherwise, the path-following methods are slow,
which is often the case for large scale problems. Other recent
developed computational methods include coordinate-wise descent
methods \cite{Pathwise}, fixed-point continuation method \cite{FPC},
sequential subspace optimization methods \cite{Sequential}, bound
optimization methods \cite{Bound}, iterated shrinkage methods
\cite{Iterative}, gradient methods \cite{Gradient}, gradient
projection for sparse reconstruction algorithm (GPSR) \cite{GP},
sparse reconstruction by separable approximation (SpaRSA)
\cite{SpaRSA} and Bregman iterative method \cite{Bregman1,Bregman2}.
Some of these methods, such as the GPSR, SpaRSA and Bregman
iterative method, can efficiently handle large-scale problems.

Besides $l_1$ norm, another typical function to represent sparsity
is $l_p$ norm ($0<p<1$). The problem is a non-convex one, thus it
is often transferred to a solvable convex problem. Typical methods
include FOCal Under-determined System Solver (FOCUSS)
\cite{FOCUSS} and Iteratively Reweighted Least Square (IRLS)
\cite{IRLS},\cite{IRLS1}. Compared with the $l_1$ norm based
methods, these methods always need more computational time.

\textbf{Greedy pursuits:}  Rather than minimize an objective
function globally, these methods make a local optimal choice after
building up an approximation at each step. Matching Pursuit (MP) and
Orthogonal Matching Pursuit (OMP)\cite{OMP1,OMP} are two of the
earliest greedy pursuit methods, then came Stagewise OMP (StOMP)
\cite{StOMP} and Regularized OMP \cite{ROMP} as their improved
versions. The reconstruction complexity of these algorithms is
around $\mathcal{O}(KMN)$, which is significantly lower than BP
methods. However, they require more measurements for perfect
reconstruction and may fail to find the sparsest solution in certain
scenarios where $l_1$ minimization succeeds. More recently, Subspace
Pursuit (SP) \cite{SP}, Compressive Sampling Matching Pursuit
(CoSaMP) \cite{CoSaMP} and Iterative Hard Thresholding method (IHT)
\cite{IHT} have been proposed by incorporating the idea of
backtracking. Theoretically they offer comparable reconstruction
quality and low reconstruction complexity as that of LP methods.
However, all of them assume that the sparsity parameter $K$ is
known, whereas $K$ may not be available in many practical
applications. In addition, all greedy algorithms are more demanding
in memory requirement.

\subsection{Our Work}
The convex optimization methods, such as $l1\_ls$ and SpaRSA, take
all the data of $\bf A$ into account for each iteration, while the
greedy pursuits consider each column of $\bf A$ for iterations. In
this paper, the adaptive filtering framework, which uses each row of
$\bf A$ for each iteration, is applied for signal reconstruction.
Moreover, instead of $l_1$ norm, we take one of the approximations
of $l_0$ norm, which is widely used in recent contribution
\cite{HUMRI}, as the sparse constraint. The authors of \cite{HUMRI}
give several effective approximations of $l_0$ norm for Magnetic
Resonance Image (MRI) reconstruction. However, their solver of this
problem adopts the traditional fix-point method, which needs much
more computational time. Thus it is hard to implement for the large
scale problem, with which our approach can effectively deal.

 According to our best knowledge, it is the
first time that the adaptive filtering framework is employed to
solve CS reconstruction problem. In our approach, two modified
stochastic gradient-based adaptive filtering methods are introduced
for signal reconstruction purpose, and a novel and improved
reconstruction algorithm is proposed in the end.

As the adaptive filtering framework can be used to solve
under-determined equation, it can be readily accepted that CS
reconstruction problem can be seen as a problem of sparse system
identification by making some correspondence. Thus, a variant of
Least Mean Square (LMS) algorithm, $l_0$-LMS, which imposes a zero
attractor on standard LMS algorithm and has good performance in
sparse system identification, is introduced to CS signal
reconstruction. In order to get better performance, an algorithm
$l_0$-Exponentially Forgetting Window LMS ($l_0$-EFWLMS) is also
adopted. The convergence of the above two methods may be slow since
$l_2$ norm and $l_0$ norm need to be balanced in their cost
functions. As regard to faster convergence, a new method named
$l_0$-Zero Attraction Projection ($l_0$-ZAP) with little sacrifice
in accuracy is further proposed. Simulations show that $l_0$-LMS,
$l_0$-EFWLMS and $l_0$-ZAP have better performances in solving CS
problem than the other typical algorithms.

The remainder of this paper is organized as follows. In Section II,
the adaptive filtering framework is reviewed and the methodological
similarity between sparse system identification and CS problem is
demonstrated. Then $l_0$-LMS, $l_0$-EFWLMS and $l_0$-ZAP are
introduced. The convergence performance of $l_0$-LMS is analyzed in
Section III. In Section IV, five experiments demonstrate the
performances of the three methods in various aspects. Finally, our
conclusion is made in Section V.

\section{Our Algorithms}
\subsection{Adaptive filtering framework to solve CS problem}
Adaptive filtering algorithms have been widely used nowadays when
the exact nature of a system is unknown or its characteristics are
time-varying. The estimation error of the adaptive filter output
with respect to the desired signal $d(n)$ is denoted by
\begin{equation}\label{lmse}
    e(n)=d(n)-{\bf{x}}^{\rm T}(n){\bf{w}}(n),
\end{equation}
where ${\bf{w}}(n)=\left[w_0(n),w_1(n),\ldots,w_{L-1}(n)\right]^{\rm
T}$ and ${\bf{x}}(n)=\left[x(n),x(n-1),\ldots,x(n-L+1)\right]^{\rm
T}$ denote the filter coefficient vector and input vector,
respectively, $n$ is the time instant, and $L$ is the filter length.
By minimizing the cost function, the parameters of the unknown
system can be identified iteratively.

Recalling the CS problem, one of its requirements is to solve the
under-determined equations $\bf y=As$. Suppose that
\begin{eqnarray}
  {\bf A} &=& \left[{{\bf a}_1^{\rm T}},{{\bf a}_2^{\rm T}},\ldots,{{\bf
a}_M^{\rm T}}\right]^{\rm T}; \\
  {\bf a}_k &=& [a_{k1},a_{k2},\ldots,a_{kN}],~k=1,2,\ldots,M; \\
  {\bf s}&=&[s_1,s_2,\ldots,s_N]^{\rm T}; \\
  {\bf y} &=& [y_1,y_2,\ldots,y_M]^{\rm T}.
\end{eqnarray}
CS reconstruction problem can be regarded as an adaptive system
identification problem by the correspondences listed in TABLE
\ref{table1}. Thus equation (\ref{Sampling}) can be solved in the
framework of adaptive filter.
\begin{table}
\renewcommand{\arraystretch}{1.3}
\caption{The correspondences between the variables in adaptive
filter and those in CS problem.} \label{table1} \centering
\begin{tabular}{cc}
\hline  adaptive filter & CS problem \\
\hline
 ${\bf x}(n)$ & ${\bf a}_k$ \\
 ${\bf w}(n)$ & ${\bf s}$ \\
  $d(n)$ & $y_k$ \\
\hline
\end{tabular} \\
\end{table}

\begin{figure}
    \centering
    \includegraphics[width=3.5in]{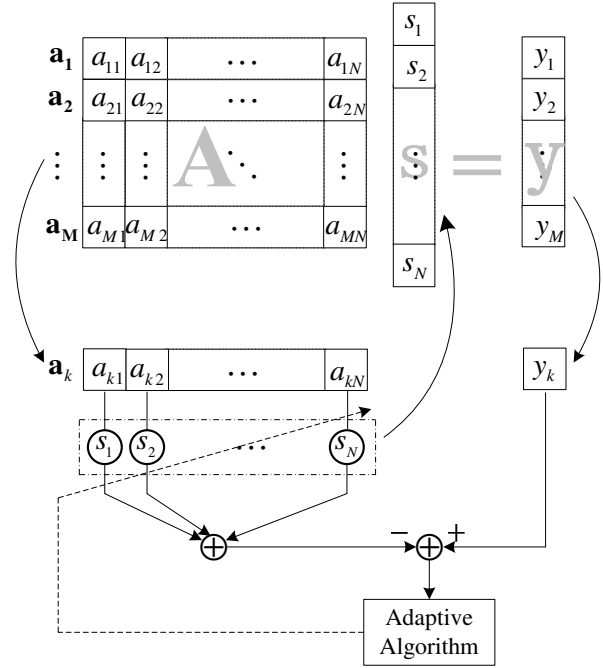}
    \caption{The framework of adaptive filter to solve CS reconstruction
problem.} \label{lmscs}
    \end{figure}

When the above adaptive filtering framework is applied to solve CS
problem, there may not be enough data to train the filter
coefficients into convergence. Thus, the rows of $\textbf{A}$ and
the corresponding elements of $\textbf{y}$ are utilized recursively.
The procedures using adaptive filtering framework are illustrated in
Fig.\ref{lmscs}. Suppose that ${\bf s}(n)$ is the updating vector,
the detailed update procedures are as follows.
\begin{enumerate}
  \item Initialize $n=1$, ${\bf s}(0)={\bf 0}$.
  \item Send data ${\bf a}_{k}$ and $y_{k}$ to adaptive filter,
  where
    \begin{equation}\label{itera}
        k=\mod(n,M)+1.
    \end{equation}
  \item Use adaptive algorithm to update ${\bf s}(n)$.
  \item Judge whether stop condition is satisfied,
    \begin{equation}\label{stop}
        \|\textbf{s}(n)-\textbf{s}(n-1)\|_2<\varepsilon  ~~~~ {\rm
        or}
        ~~~~n>C,
    \end{equation}
    where $\varepsilon>0$ is a given error tolerance and $C$ is a given
    maximum iteration number.
  \item When satisfied, send ${\bf s}(n)$ back to $\bf s$ and exit;
  otherwise $n$ increases by one and go back to 2).
\end{enumerate}

Adaptive filtering methods are well-known while CS is a popular
topic in recent years, so it is surprising that no literature
employs adaptive filtering structure in CS reconstruction problem.
The reason might be that the aim of CS is to reconstruct a sparse
signal while the solutions to general adaptive filtering algorithms
are not sparse. In fact, several LMS variations
\cite{l0lms,IPNLMS,Exploiting}, with some sparse constraints added
in their cost functions, exist in sparse system identification.
Thus, these methods can be applied to solve CS problem.

In following subsections, $l_0$-LMS algorithm and the idea of zero
attraction will be firstly introduced. Then $l_0$-EFWLMS, which
imposes zero attraction on EFW-LMS, is introduced for better
performance. Finally, to speed up the convergence of the two new
methods, a novel algorithm $l_0$-ZAP, which adopts zero attraction
in solution space, is further proposed.

\subsection{Based on $l_0$-LMS algorithm}
LMS is the most attractive one in all adaptive filtering algorithms
because of its simplicity, robustness and low computation cost. In
traditional LMS the cost function is defined as squared error,
\begin{equation}\label{lmsxi}
    \xi_{\rm LMS}(n)=|e(n)|^2.
\end{equation}
Consequently, the gradient descent recursion of the filter
coefficient vector is
\begin{equation}\label{lmsgdr}
    {\bf w}(n+1)={\bf w}(n)+\mu e(n){\bf x}(n),
\end{equation}
where positive parameter $\mu$ is called step-size.

In order to improve the convergence performance when the unknown
parameters are sparse, a new algorithm $l_0$-LMS \cite{l0lms} is
proposed by introducing a $l_0$ norm penalty to the cost function.
The new cost function is defined as
\begin{equation}
\xi_{\rm new}(n) = |e(n)|^2 + \gamma\|{\bf w}(n)\|_0,
\label{l0LMScost}
\end{equation}
where $\gamma>0$ is a factor to balance the new penalty and the
estimation error. Considering that $l_0$ norm minimization is an NP
hard problem, $l_0$ norm is generally approximated by a continuous
function. A popular approximation \cite{Weston} is
\begin{equation}
\|{\bf w}(n)\|_0 \approx \sum_{i=0}^{L-1}\left( 1-{\rm
e}^{-\alpha|w_i(n)|} \right), \label{l0app}
\end{equation}
where the two sides of (\ref{l0app}) are strictly equal when
parameter $\alpha$ approaches infinity. According to (\ref{l0app}),
the proposed cost function can be rewritten as
\begin{equation}
\xi_{l_0-{\rm LMS}}(n) = |e(n)|^2 + \gamma\sum_{i=0}^{L-1}\left(
1-{\rm e}^{-\alpha|w_i(n)|} \right). \label{l0LMScost1}
\end{equation}

By minimizing (\ref{l0LMScost1}), the new gradient descent recursion
of filter coefficients is
\begin{equation}
w_i(n+1) = w_i(n) + \mu e(n)x(n-i) -
\kappa\alpha \text{sgn}(w_i(n)) {\rm e}^{-\alpha|w_i(n)|}, \quad
\forall 0\leq i < L,
\label{l0LMSrecu}
\end{equation}
where $\kappa=\mu\gamma$ and sgn($\cdot$) is a component-wise sign
function defined as
\begin{equation}\label{sgnfunc}
    {\text{sgn}}(x)=\left\{
    \begin{array}{cl} \frac{\textstyle x}{\textstyle |x|} & x\neq 0; \\
    0 & {\rm elsewhere}. \end{array} \right.
\end{equation}

To reduce the computational complexity of (\ref{l0LMSrecu}),
especially that caused by the last term, the first order Taylor
series expansion of exponential functions is taken into
consideration,
\begin{equation}\label{taylor}
    {\rm e}^{-\alpha |x|}\approx \left\{
    \begin{array}{cl} 1-\alpha|x| & |x|\le\frac{\textstyle 1}{\textstyle \alpha}; \\
    0 & {\rm elsewhere}. \end{array} \right.
\end{equation}
Note that the approximation of (\ref{taylor}) is bound to be
positive because the value of exponential function is larger than
zero. Thus the final gradient descent recursion of filter
coefficient vector is
\begin{equation}
{\bf w}(n+1) = {\bf w}(n) + \mu e(n){\bf x}(n) + \kappa {\bf g}({\bf
w}(n)), \label{l0LMSrecu1}
\end{equation}
 where
\begin{equation}\label{gxfunction}
    {\bf g}({\bf w}(n)) = \left[ g(w_0(n)),g(w_1(n)),\ldots,g(w_{L-1}(n)) \right]^{\rm T}
\end{equation}
and
\begin{equation}\label{falpha}
    g(x)=\left\{
    \begin{array}{cc} \alpha^2x+\alpha & -\frac{\textstyle 1}{\textstyle \alpha}\le x < 0; \\
    \alpha^2x-\alpha & 0 < x \le \frac{\textstyle 1}{\textstyle \alpha}; \\
    0 & {\rm elsewhere}. \end{array} \right.
\end{equation}

The last term of (\ref{l0LMSrecu1}) is called \emph{zero attraction}
term, which imposes an attraction to zero on small coefficients.
Since zero coefficients are the majority in sparse systems, the
convergence acceleration of zero coefficients will improve
identification performance. In CS, the zero attraction term will
ensure the sparsity of the solution.

By utilizing the correspondence in TABLE \ref{table1}, the final
solution to CS problem can be obtained, which is summarize as Method 1.

\begin{table}[t]
\renewcommand{\arraystretch}{1.3}
\label{algorithm1}
\centering
\begin{tabular}{l}
\hline  Method 1. $l_0$-LMS method for CS \\
\hline
1: Initialize ${\bf s}(0)={\bf 0}$, $n$=1, choose $\mu,\alpha,\kappa$;\\
2: while stop condition (\ref{stop}) is not satisfied; \\
3: ~~~~~~Determine the input vector ${\bf x}(n)$ and desired signal $d$(n) \\
   ~~~~~~~~~~~~~~~~~~$k$ = mod($n,M$)+1;\\
   ~~~~~~~~~~~~~~~~~~${\bf x}(n)$ = ${\bf a}_k$;\\
   ~~~~~~~~~~~~~~~~~~$d(n)$ = $y_k$;\\
4: ~~~~~~Calculate error $e(n)$\\
   ~~~~~~~~~~~~~~~~~~$e(n)$ = $d(n)-{\bf x}^{\rm T}(n){\bf s}(n)$;\\
5: ~~~~~~Update ${\bf s}(n)$ using LMS \\
   ~~~~~~~~~~~~~~~~~~${\bf s}(n)$ = ${\bf s}(n-1)+\mu e(n){\bf x}(n)$;\\
6: ~~~~~~Impose a zero attraction \\
   ~~~~~~~~~~~~~~~~~~${\bf s}(n)$ = ${\bf s}(n)+\kappa {\bf g}\left({\bf s}(n-1)\right)$;\\
7: ~~~~~~Iteration number increases by one \\
   ~~~~~~~~~~~~~~~~~~$n=n+1;$\\
8: End while. \\
 \hline
\end{tabular} \\
\end{table}

\subsection{Based on $l_0$-EFWLMS algorithm}
Recursive Least Square (RLS) is another popular adaptive filtering
algorithm \cite{Cowan}, \cite{Haykin}, whose cost function is
defined as the weighted sum of continuous squared error sequence,
\begin{equation}\label{rlsxi}
    \xi_{\rm RLS}(n)=\sum_{i=1}^{n}\lambda^{n-i} |e(i)|^2.
\end{equation}
where $0\ll\lambda<1$ is called forgetting factor and
\begin{equation}\label{ei}
    e(i)=d(i)-{\bf x}^{\rm T}(i){\bf w}(n).
\end{equation}

The RLS algorithm is difficult to implement in CS because it costs a
lot of computing resources. However, motivated by RLS, the
approximation of its cost function with shorter sliding-window is
considered, which suggests a new penalty
\begin{equation}\label{rlsxi1}
    \xi_{\rm EFW-LMS}(n)=\sum_{i=n-Q+1}^{n}\lambda^{n-i} |e(i)|^2,
\end{equation}
where $Q$ is the length of the sliding-window. The algorithm, which
minimizes (\ref{rlsxi1}), is called Exponentially Forgetting Window
LMS (EFW-LMS) \cite{EFWLMS}. The gradient descent recursion of the
filter coefficient vector is
\begin{equation}\label{l0efwlms}
    {\bf w}(n+1)={\bf w}(n)+\mu {\bf X}(n){\bf \Lambda}{\bf e'}(n),
\end{equation}
where
\begin{align}
    {\bf X}(n)&=\left[{\bf x}(n-Q+1),{\bf x}(n-Q+2),\ldots, {\bf x}(n)
    \right],\label{Xn}\\
    {\bf \Lambda}&=\left[ \begin{array}{cccc}
      \lambda^{Q-1} & 0 & \ldots & 0\\
      0 & \lambda^{Q-2} & \ldots & 0\\
      \vdots & \vdots & \ddots & \vdots \\
      0 & 0 & \ldots & 1
    \end{array}
    \right],\label{lambda}\\
    {\bf e'}(n)&=\left[e(n-Q+1),e(n-Q+2),\ldots, e(n)
    \right]^{\rm T} \nonumber \\
    &={\bf d'}(n)-{\bf X}^{\rm T}(n){\bf w}(n),\label{e'}
\end{align}
and
\begin{equation}\label{dn}
    {\bf d'}(n)=\left[d(n-Q+1),d(n-Q+2),\ldots, d(n)
    \right]^{\rm T}.
\end{equation}

In order to obtain sparse solutions in CS problem, zero attraction
is employed again. Thereby the final gradient descent recursion of
the filter coefficient vector is
\begin{equation}\label{l0efwlms}
    {\bf w}(n+1)={\bf w}(n)+\mu {\bf X}(n){\bf \Lambda}{\bf e'}(n)+\kappa {\bf g}({\bf
w}(n)).
\end{equation}
This algorithm is denoted as $l_0$-EFWLMS.

The method to solve CS problem utilizing the correspondence in TABLE
\ref{table1} based on $l_0$-EFWLMS is summarized in Method 2.
\begin{table}[t]
\renewcommand{\arraystretch}{1.3}
\label{algorithm1}
\centering
\begin{tabular}{l}
\hline  Method 2. $l_0$-EFWLMS method for CS \\
\hline
1: Initialize ${\bf s}(0)=0$, choose $Q,\mu,\lambda,\alpha,\kappa$;\\
2: while stop condition (\ref{stop}) is not satisfied; \\
3: ~~~~~~Determine $Q$ input vectors ${\bf x}(n-Q+1),\cdots,{\bf x}(n)$\\
   ~~~~~~~~~and $Q$ desired signals $d(n-Q+1),\cdots,d(n)$ \\
   ~~~~~~~~~~~~~~~~~~For $i=n-Q+1,...,n$\\
   ~~~~~~~~~~~~~~~~~~~~~~~~~~~$k$ = mod($i,M$)+1;\\
   ~~~~~~~~~~~~~~~~~~~~~~~~~~~${\bf x}(i)$ = ${\bf a}_k$;\\
   ~~~~~~~~~~~~~~~~~~~~~~~~~~~$d(i)$ = $y_k$;\\
   ~~~~~~~~~~~~~~~~~~End for; \\
4: ~~~~~~Calculate error vector ${\bf e'}(n)$\\
   ~~~~~~~~~~~~~~~~~~${\bf e'}(n)$ = ${\bf d'}(n)-{\bf X}^{\rm T}{\bf s}(n-1)$;\\
5: ~~~~~~Update ${\bf s}(n)$ using EFW-LMS \\
   ~~~~~~~~~~~~~~~~~~${\bf s}(n)$ = ${\bf s}(n-1)+\mu {\bf X}(n){\bf \Lambda}{\bf e'}(n)$;\\
6: ~~~~~~Impose a zero attraction \\
   ~~~~~~~~~~~~~~~~~~${\bf s}(n)$ = ${\bf s}(n)+\kappa {\bf g}\left({\bf s}(n-1)\right)$;\\
7: ~~~~~~Iteration number increases by one \\
   ~~~~~~~~~~~~~~~~~~$n=n+1$;\\
8: End while. \\
 \hline
\end{tabular} \\
\end{table}

\subsection{Based on $l_0$-ZAP algorithm}
\begin{figure*}[t]
    \centering
    \includegraphics[width=15cm]{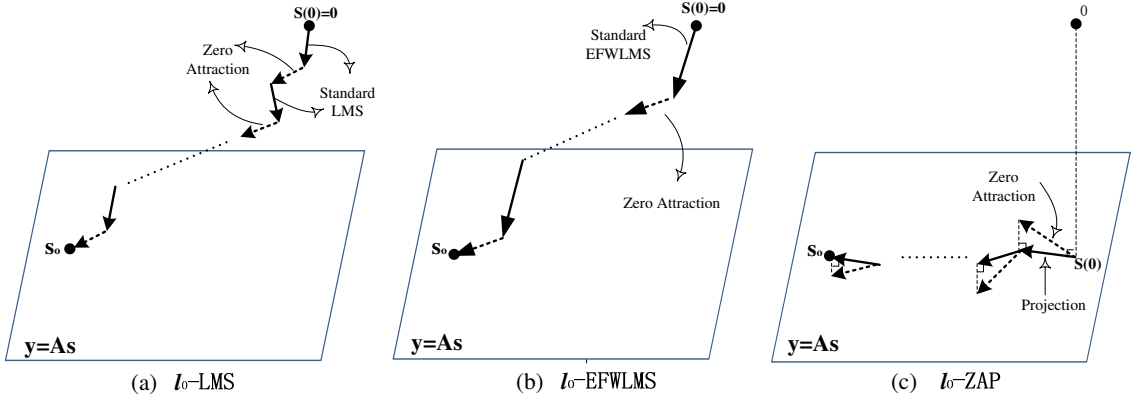}
    \caption{The updating procedures of the three methods, where $\bf s_o$ denotes the original signal and ${\bf s}(0)$ denotes the initial value.(a) $l_0$-LMS; (b) $l_0$-EFWLMS; (c) $l_0$-ZAP.} \label{relation}
    \end{figure*}

The two methods described above $l_0$-LMS and $l_0$-EFWLMS can be
considered as solutions to $l_2-l_0$ problem. Observing
(\ref{l0LMSrecu1}) and (\ref{l0efwlms}), it is obvious that both
gradient descent recursions are consisted of two parts.
\begin{equation}
{\bf w}_{\rm new} = {\bf w}_{\rm prev} + {\rm gradient~correction} +
{\rm zero~attraction},
\end{equation}
The gradient correction term is to ensure $\bf y=As$, and the zero
attraction term is to guarantee the sparsity of the solution. Taking
both parts into account, the sparse solution can finally be
extracted. The updating procedures of the two methods proposed are
shown in Fig.\ref{relation}.(a) and Fig.\ref{relation}.(b). However,
convergence of the recursions may be slow because the two parts are
hard to balance.

According to the discussions above, CS problem (\ref{Sampling}) is
ill-conditioned and its solution is a $N-M$ subspace. It implies
that the sparse solution can be searched iteratively in the solution
space in order to speed up convergence. That is, the gradient
correction term can be omitted. The updating procedures are
demonstrated in Fig.\ref{relation}.(c), where the initial vector of
${\bf s}(0)$ is taken as the Least Square (LS) solution, which
belongs to the solution space. Then in iterations, only the zero
attraction term is used for updating the vector. The updated vector
is replaced by the projection of the vector on solution space as
soon as it departs from the solution space. Particularly, suppose
${\bf s}(n)$ is the result gained after $n$th zero attraction, its
projection vector in the solution space satisfy the following
equation
\begin{equation}\label{l2project}
    \hat{\bf s}(n)=\arg \min_{{\bf s}'(n)} \|{\bf s}'(n)-{\bf s}(n)\|_2^2,~~~~ {\rm
    s.t.}~~~{\bf A}{\bf s}'(n)={\bf y}.
\end{equation}
Laplacian Method can be used to solve (\ref{l2project}),
\begin{equation}\label{projectsolution}
    \hat{\bf s}(n)
    ={\bf s}(n)+{\bf A}^+{({\bf y}-{\bf A}{\bf s}(n))},
\end{equation}
where ${\bf A}^+ = {\bf A}^{\rm T}{({\bf AA}^{\rm T})}^{-1}$ is the
Pseudo-inverse matrix of Least Square. This method is called
$l_0$-Zero Attraction Projection ($l_0$-ZAP), which is summarized in
Method 3.

\begin{table}[t]
\renewcommand{\arraystretch}{1.3}
\label{algorithm1}
\centering
\begin{tabular}{l}
\hline  Method 3. $l_0$-ZAP method for CS \\
\hline
1: Initialize ${\bf s}(0)={\bf {A^+}y}$, choose $\alpha,\kappa$,\\
2: while stop condition (\ref{stop}) is not satisfied \\
3: ~~~~Update ${\bf s}(n)$ using zero attraction \\
   ~~~~~~~~~~~${\bf s}(n)$ = $ {\bf s}(n-1)+\kappa {\bf g}({\bf s}(n-1))$;\\
4: ~~~~Project ${\bf s}(n)$ on the solution space \\
   ~~~~~~~~~~~${\bf s}(n)$ = ${\bf s}(n)+{{\bf A}^+}{({\bf y}-{\bf As}(n))}$;\\
5: ~~~~Iteration number increases by one \\
   ~~~~~~~~~~~$n=n+1$;\\
6: End while \\
 \hline
\end{tabular} \\
\end{table}

\subsection{Discussion}
The typical performance of the three proposed methods are briefly
discussed here.
\begin{itemize}
  \item Memory requirement: $l_0$-LMS and $l_0$-EFWLMS need storage for $\bf y,~A$, and $\bf
  s$, so both their storage requirements are about $MN+M+N$. $l_0$-ZAP
needs additional storage for, at least, the Pseudoinverse matrix of
Least Square, ${\bf A}^+$. For large scale situation, $l_0$-ZAP
requires about twice the memory of the other two algorithms.

  \item Computational complexity: the total computational complexity depends on
the number of iterations required and the complexity of each
iteration.
  First, the complexity of each iteration of these methods will be analyzed.
For simplicity, the complexity of each period, instead of that of
each iteration, will be
  discussed.
  Here, a period is defined as all data in matrix $\bf A$ has been used for one
  time. For example, in one period, (\ref{l0LMSrecu1}) is iterated $M$ times in $l_0$-LMS
  and the projection is used once in $l_0$-ZAP. For each period, the
  complexity of the three methods is listed in TABLE
  \ref{complexity}. It can be seen that
  \begin{equation}\label{eachperiod}
    l_0\textrm{-ZAP} < l_0\textrm{-LMS} < l_0\textrm{-EFWLMS}.
  \end{equation}

\begin{table}[!t]
\renewcommand{\arraystretch}{1.3}
\caption{The computational complexity of different method in each
period.} \label{complexity} \centering
\begin{tabular}{cccccc}
\hline  Methods & Multiplications & Additions & Times of Zero Attraction \footnotemark{} \\
\hline
$l_0$-LMS & $3MN$  &  $2MN$   & $M$ \\
\hline
$l_0$-EFWLMS & $(2Q+1)MN$  &  $(2Q+1)MN$   & $M$ \\
\hline
$l_0$-ZAP & $2MN$  &  $2MN+N+M$   & $1$ \\
\hline
\end{tabular} \\
\begin{tabular}{c} [1]Please note the computations of zero attraction is not included in
the\\

above multiplicaitons and additions.
\end{tabular}
\end{table}

Second,  the number of periods of these methods will be discussed.
It is impossible to accurately
  predict the number of periods of the three proposed methods
  required to find an approximate solution. However, according to
  the above discussion, the following equation is always satisfied for
  the number of periods
  \begin{equation}\label{numberperiod}
    l_0\textrm{-ZAP} < l_0\textrm{-EFWLMS} < l_0\textrm{-LMS}.
  \end{equation}

Thus, taking both (\ref{eachperiod}) and (\ref{numberperiod}) into
consideration, $l_0$-ZAP has significantly lower computation
complexity than $l_0$-LMS and $l_0$-EFWLMS. Because $l_0$-LMS has
lower complexity for each period but larger number of periods than
$l_0$-EFWLMS, a comparison between $l_0$-LMS and $l_0$-EFWLMS is
hard to make.

  \item De-noise performance: $l_0$-LMS and $l_0$-EFWLMS inherit the merit of LMS algorithm that
has good de-noise performance. For $l_0$-ZAP,
\begin{equation}\label{denoise}
    \bf y=As+v=A(s+\hat{v})=A\hat{s}
\end{equation}
where $\bf \hat{v}=A^+{v}$ and $\bf v$ is an additive noise. Thus,
the iterative vector is not projected on the true solution set $\bf
s$ but the solution space $\bf{\hat{s}}$ with additive noise $\bf
\hat{v}$. However, we have
\begin{equation}\label{noiserelation}
    {\rm E}\left\{ {\bf \hat{v}}^{\rm T}{\bf \hat{v}} \right\}
    \approx \frac{M}{N}{\rm E}\left\{ {\bf v}^{\rm T}{\bf v}
    \right\},
\end{equation}
where ${\rm E}(\cdot)$ denotes the expectation. The proof of
(\ref{noiserelation}) is in Appendix A. Equation
(\ref{noiserelation}) shows that the power of $\bf \hat{v}$ is far
smaller than that of $\bf v$ since $M\ll N$. Moreover, the dimension
of $\bf v$ (e.g. $M$) is far smaller than that of $\bf \hat{v}$
(e.g. $N$). Therefore, $l_0$-ZAP also has good de-noise performance.

  \item Implementation difficulty: $l_0$-ZAP need two parameters
$\alpha$ and $\kappa$, while in $l_0$-LMS and $l_0$-EFWLMS, there is
another parameter $\mu$ to be chosen. Thus, $l_0$-ZAP is easier to
control than the other two algorithms.
\end{itemize}

\subsection{Some Comments}
\textbf{Comment 1:} Besides the proposed $l_0$-LMS and $l_0$-EFWLMS,
the idea of zero attraction can be readily adopted to improve most
LMS variants, e.g. Normalized LMS (NLMS), which may be more
attractive than LMS because of its robustness. The gradient descent
recursion of the filter coefficient vector of $l_0$-NLMS is
\begin{equation}
{\bf w}(n+1) = {\bf w}(n) + \mu \frac{e(n){\bf x}(n)}{\beta+{\bf
x}^{\rm T}(n){\bf x}(n)} + \kappa {\bf g}\left({\bf w}(n)\right),
\label{l0NLMSrecu}
\end{equation}
where $\beta>0$ is the regularization parameter. These variants can
also improve the performance in sparse signal reconstruction.

\textbf{Comment 2:} Equation (\ref{l0app}) is one of the multiple
approximations of $l_0$ norm. In fact, many other continuous
functions can be used for zero attraction. For example, an
approximation suggested by Weston et al. \cite{Weston} is
\begin{equation}\label{logapp}
    \|{\bf w}\|_0 \approx \sum_{i=0}^{L-1}
    \frac{|w_i|}{|w_i|+\delta},
\end{equation}
where $\delta$ is a small positive number. By minimizing
(\ref{logapp}), the corresponding zero attraction is
\begin{equation}\label{gxfunction1}
    \kappa {\bf g} ({\bf w}) = \kappa\left[g(w_0),g(w_1),\ldots,g(w_{L-1})\right]^{\rm
    T},
\end{equation}
where
\begin{equation}\label{falpha1}
    g(x)=\frac{\delta \text{sgn}(x)}{(|x|+\delta)^2}.
\end{equation}
This zero attraction term can also be used in the proposed
$l_0$-LMS, $l_0$-EFWLMS and $l_0$-ZAP.

\section{Convergence analysis}
In this section, we will analyse the convergence performance of
$l_0$-LMS. The steady-state mean square derivation between the
original signal and the reconstruction signal will be analyzed and
the bound of parameter $\mu$ to guarantee convergence will be
deduced.

  \emph{Theorem 1:}Suppose that ${\bf s}$ is the original signal, and $\hat{\bf s}$ is the
  reconstruction signal by $l_0$-LMS, the final mean
  square
  derivation in steady state is
  \begin{equation}\label{theorem1}
  {\rm E}\left\{\|\hat{\bf s}-{\bf{s}}\|_2^2\right\} =C\left[2\kappa (1-\frac{\mu}{M})a+\kappa^2 b + \frac{N\mu^2}{M}
    P_0\right],
  \end{equation}
  where
\begin{eqnarray}\label{e9}
    C&=&\frac{M^2}{2 \mu M-(N+2)\mu^2}; \\
    a &=& {\rm E}\left\{(\hat{\bf s}-{\bf{s}})^{\rm T}{\bf g}(\hat{\bf s})\right\}; \\
    b &=& {\rm E}\left\{{\bf g}^{\rm T}(\hat{\bf s}){\bf g}(\hat{\bf s})\right\}; \\
    P_0&=&{\rm E}\left\{v^2\right\},
\end{eqnarray}
$P_0$ is the power of measurement noise. At the same time, in order
to guarantee convergence, parameter $\mu$ should satisfy
\begin{equation}\label{mucond}
    0<\mu<\frac{2M}{N+2}.
\end{equation}


The proof of the theorem is postponed to Appendix B.

As shown in Theorem 1, the final derivation is proportional to
$\kappa$ and the power of measurement noise. Thus a large $\kappa$
will result in a large derivation; However, a small $\kappa$ means a
weak zero attraction that will induce a slower convergence.
Therefore, the parameter $\kappa$ is determined by a trade-off
between convergence rate and reconstruction quality in particular
applications.

By equation (\ref{alphabound}) and (\ref{betabound}) in appendix we
have the following corollary

 \emph{Corollary 1:}The upper bound of derivation is
  \begin{equation}\label{theorem1}
  {\rm E}\left\{\|\hat{\bf s}- {\bf{s}}\|_2^2\right\}\leq
   C\left[2\kappa (1-\frac{\mu}{M})(N+\alpha \|{\bf s}\|_1)+N\kappa^2\alpha^2 + \frac{N\mu^2}{M}
    P_0\right].
  \end{equation}

The upper bound is a constant under a given signal, thus it can be
regarded as a rough criterion to choose the parameters.

\section{Experiment Results}
The performances of the presented three methods are experimentally
verified and compared with typical CS reconstruction algorithms
BP\cite{CSbase}, SpaRSA\cite{SpaRSA}, GPSR-BB\cite{GP},
$l1\_ls$\cite{l1ls}, Bregman iterative algorithm based on FPC
(FPC\_AS)\cite{Bregman2}, IRLS\cite{IRLS} and OMP\cite{OMP}. In the
following experiments, these algorithms are tested with parameters
recommended by respective authors. The entries of $M\times N$
sensing matrix $\bf A$ are independently generated from normal
distribution with mean zero and variance $1/M$. The locations of $K$
nonzero coefficients of sparse signal $\bf{s}$ are randomly chosen
with uniform distribution $[1,N]$. The corresponding nonzero
coefficients are Gaussian with mean zero and unit variance. Finally
the sparse signal is normalized. The measurements are generated by
the following noisy model
\begin{equation}\label{noisymodel}
  \bf y = As + v,
\end{equation}
where $\bf v$ is an additive white Gaussian noise with covariance
matrix ${\sigma}^2{\bf I}_M$ (${\bf I}_M$ is an $M \times M$
identity matrix).

The parameters in stop condition (\ref{stop}) are
$\varepsilon=10^{-4}$ for all three methods, $C=10^5$ for $l_0$-LMS
and $l_0$-EFWLMS, $C=10^3$ for $l_0$-ZAP.

    \emph{Experiment 1. Algorithm Performance:}
    In this experiment, the performances of the three
    proposed methods in solving CS problem are tested. The parameters used for the
    signal model (\ref{noisymodel}) are $\sigma=3.2\times 10^{-3},N=1000,M=200,K=30$. The parameters for
    the three methods are as follows:
    \begin{itemize}
      \item $l_0$-LMS: $\alpha=10$, $\mu=0.1,~\kappa=2\times 10^{-6}$;
      \item $l_0$-EFWLMS: $\alpha=10$, $\mu=0.1,~\kappa=2\times 10^{-6}$, $Q=4$, $\lambda=0.8$;
      \item $l_0$-ZAP: $\alpha=10$, $\kappa=5\times 10^{-4}$.
    \end{itemize}

\begin{figure}
    \centering
    \includegraphics[width=4in]{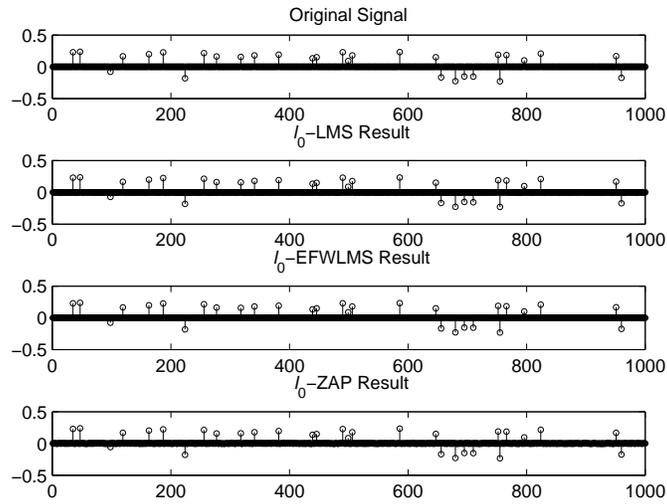}
    \caption{Reconstruction result of the three proposed methods.} \label{result}
    \end{figure}

\begin{figure}
    \centering
    \includegraphics[width=4in]{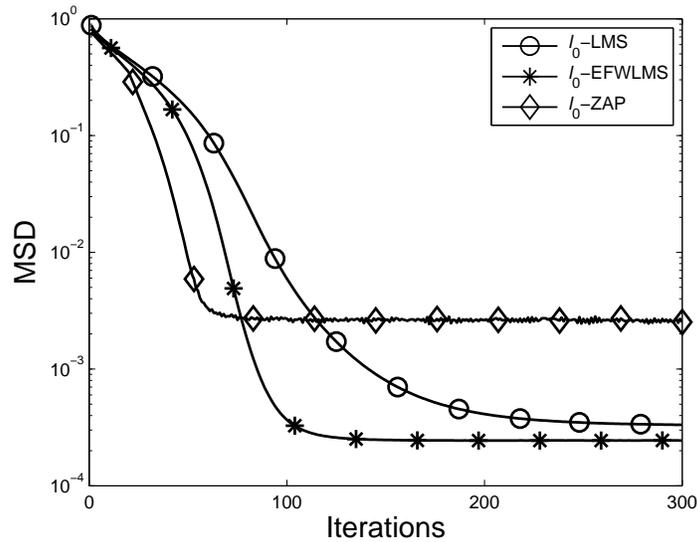}
    \caption{Convergence performances of the three proposed methods.} \label{msd}
    \end{figure}

    The original signal and the estimation
    results obtained with $l_0$-LMS, $l_0$-EFWLMS, and $l_0$-ZAP are shown in
    Fig.\ref{result}. It can be seen that
    all three proposed methods reconstruct the original signal.
    The convergence curves of the three
    methods are demonstrated in Fig.\ref{msd}, where MSD denotes Mean Square Derivation.
    For $l_0$-LMS and $l_0$-EFWLMS, all data of matrix $\bf A$ is used
    once in each iteration (Please note that the stop condition is not used
    here).
    As can be seen in Fig.\ref{msd}, $l_0$-EFWLMS has the smallest MSD after
    convergence and $l_0$-ZAP achieves the fastest convergence with sacrifice in reconstruction quality.

    To compare with the other algorithms,
    CPU time is used as an index of complexity, although it gives only a rough estimation of
    complexity. Our simulations are performed in MATLAB 7.4
    environment using an Intel T8300, 2.4GHz processor with 2GB of
    memory, and under Microsoft Windows XP operating system.
    The final average CPU time (of total $10$ times, in seconds) and MSD are listed in
    TABLE \ref{complexitylist}. Here, the parameter
    in IRLS is $p=1/2$. It can be seen that the proposed three methods  have the least MSD.
    In addition, $l_0$-ZAP is fastest among listed algorithms, though $l_0$-LMS
    and $l_0$-EFWLMS have no significant advantage over the other algorithms.

    \begin{table}[t]
    \renewcommand{\arraystretch}{1.3}
    \caption{The CPU time and MSD.}
    \label{complexitylist} \centering
    \begin{tabular}{cccccc}
    \hline  algorithms & average CPU time (in sec) & MSD \\
    \hline
    BP &  0.582 &   $1.1\times 10^{-2}$     \\
    OMP &  0.094   &   $7.18\times 10^{-2}$    \\
    IRLS &   1.836  &   $2.31\times 10^{-3}$    \\
    $l1\_ls$ &   1.436  &   $7.68\times 10^{-2}$    \\
    SpaRSA &  0.221 &   $7.25\times 10^{-2}$     \\
    GPSR-BB &  0.266 &   $7.43\times 10^{-2}$     \\
    FPC-AS &  0.086 &   $7.38\times 10^{-2}$     \\
    $l_0$-LMS & 1.152 & $3.33\times 10^{-4}$ \\
    $l_0$-EFWLMS & 1.544 & $2.44\times 10^{-4}$ \\
    $l_0$-ZAP & 0.068 & $2.25\times 10^{-3}$ \\
    \hline
    \end{tabular} \\
    \end{table}

\begin{figure}
    \centering
    \includegraphics[width=4in]{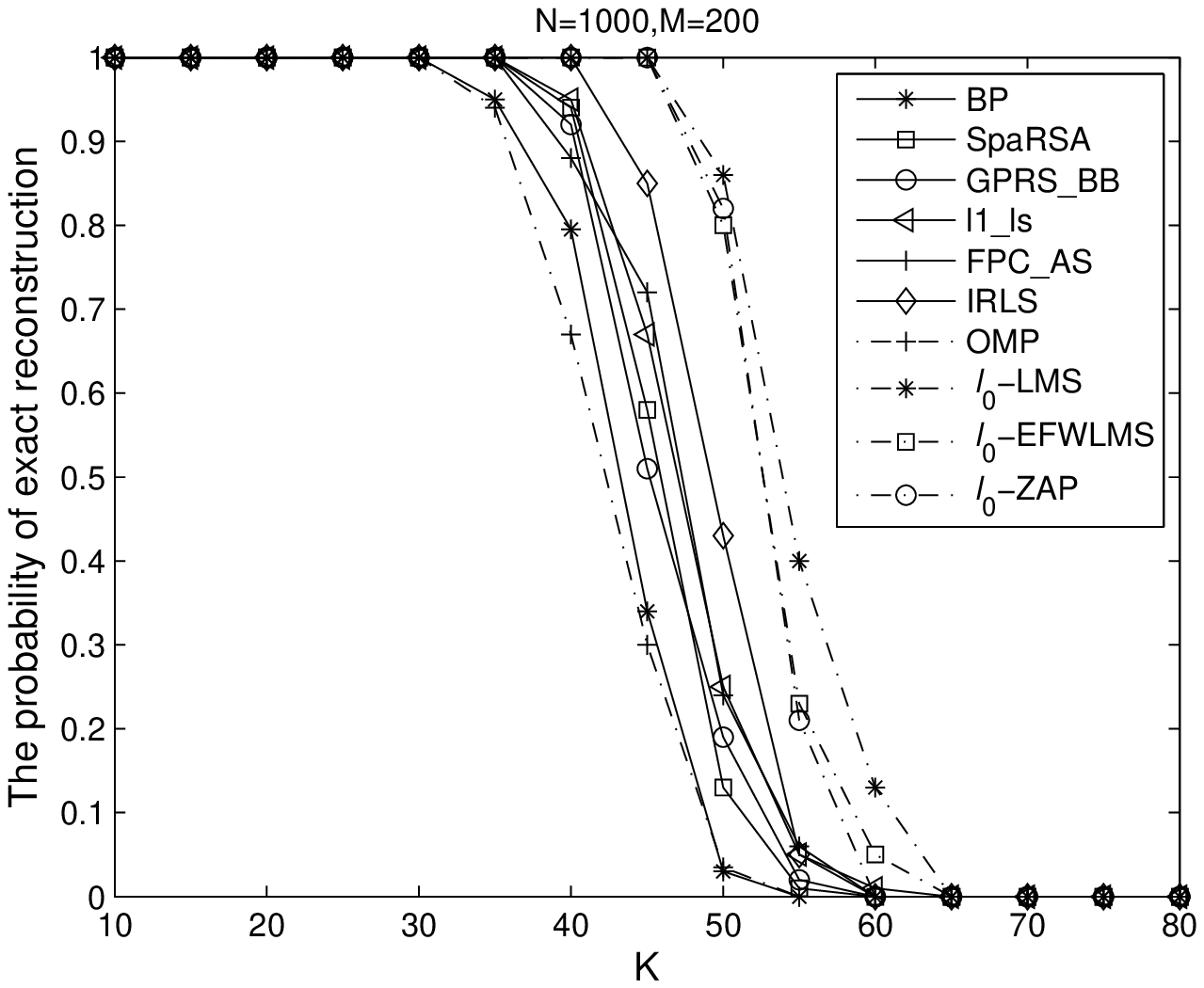}
    \caption{The probability of exact reconstruction versus sparsity $K$.} \label{vark}
    \end{figure}

    \emph{Experiment 2. Effect of Sparsity on the performance:}
    This experiment explores the answer to this question: with the proposed
    methods, how
    sparse a source vector $\bf s$ should be to make its estimation
    possible under given number of measurements.
    The parameters are the same as the first experiment except that
    the noise variance is zero. Different sparsities (i.e. $K$) are chosen
    from $10$ to $80$. For each $K$, $200$ simulations are
    conducted to calculate the probability of exact reconstruction
    in different algorithms. The results for all seven algorithms
    are demonstrated in Fig.\ref{vark}. As can be seen, performances of
    the three proposed methods far exceed those of the other algorithms. While all
    the other algorithms fail when sparsity $K$ is larger than $40$, the three methods proposed
    succeed until sparsity $K$ reaches $45$. In addition, the proposed three methods have
    similar good performances.

\begin{figure}
    \centering
    \includegraphics[width=4in]{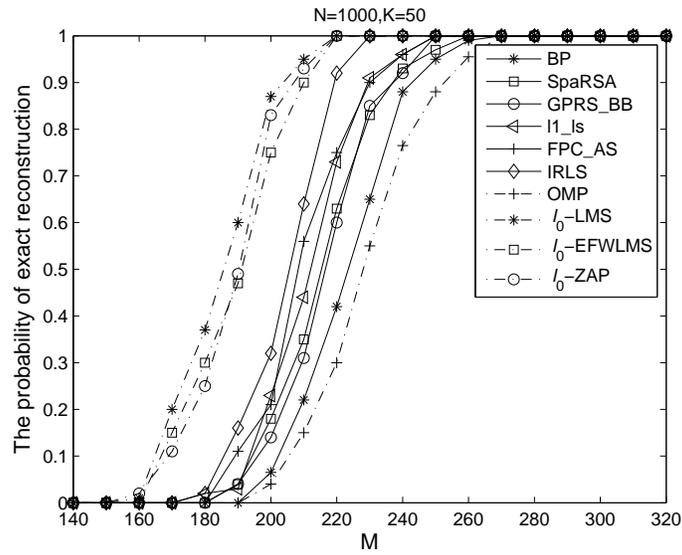}
    \caption{The probability of exact reconstruction versus measurement number $M$.} \label{varm}
    \end{figure}

    \emph{Experiment 3. Effect of number of measurements on
    the performance:}
     This experiment is to investigate the probability of exact
    recovery when
    given different numbers of measurements and a fixed signal sparsity $K=50$.
    The same setups of the first experiment is used except that the noise variance is zero.
    Different
    numbers of measurements $M$ are chosen from $140$ to $320$. All
    these algorithms are repeated $200$ times for each value of $M$,
    and the probability curves are shown in Fig.\ref{varm}.
    Again, it can be seen that the three proposed methods have the
    best performances. While all other algorithms fail when
    the measurement number $M$ is lower than $230$, the three proposed methods
    can still reconstruct exactly the original signal until $M$ reaches $220$. Meanwhile,
    the proposed algorithms have comparable good performances.

\begin{figure}
    \centering
    \includegraphics[width=4in]{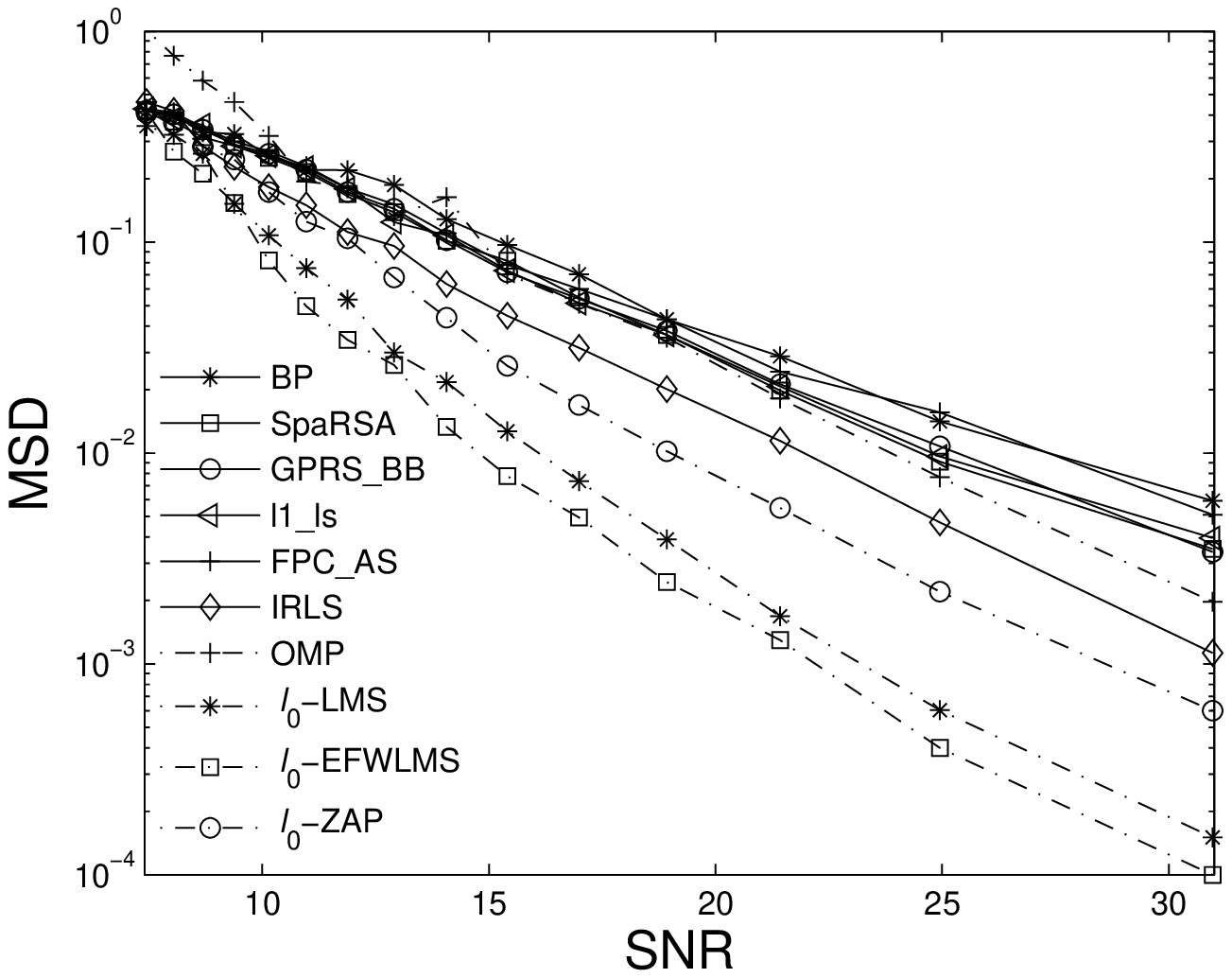}
    \caption{The reconstruction MSD versus SNR.} \label{varfig}
    \end{figure}

    \emph{Experiment 4. Robustness against noise:}
    The fourth experiment is to test the effect of signal-to-noise ratio (SNR) on
    reconstruction performance, where SNR is defined as $\text{SNR}=10\log{\|\bf As\|_2^2}/{\|\bf v\|_2^2}$. The parameters are the same as the
    first experiment and SNR is chosen from
    $4$dB to $32$dB. For each SNR, all
    these algorithms are repeated $200$ times to
    calculate the MSD. Fig.\ref{varfig} shows that the three new
    methods have better performances than the other traditional algorithms in all
    SNR. With the same SNR, the proposed algorithms can acquire small MSDs.
    In addition, the $l_0$-EFWLMS has the smallest
    MSD and $l_0$-ZAP has the largest MSD in the three new methods. Obviously, the above results
    are consistent with discussions in previous sections.

\begin{figure}
    \centering
    \includegraphics[width=4in]{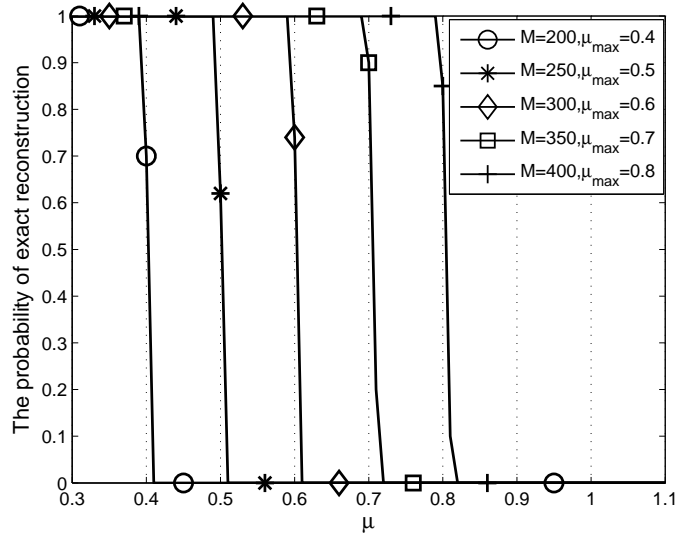}
    \caption{The probability of exact reconstruction of $l_0$-LMS versus $\mu$ with different $M$.} \label{mufig}
    \end{figure}

    \emph{Experiment 5. Effect of parameter $\mu$ on the performance of $l_0$-LMS:} In this experiment, the condition (\ref{mucond}) on step-size
    to guarantee the convergence of $l_0$-LMS will be verified. The setups of this
    experiment are the same as the first experiment except that $M=\{200,250,300,350,400\}$.
    For each $M$, $100$ simulations are
    conducted to calculate the probability of exact reconstruction using
    $l_0$-LMS with the parameters $\alpha=10$, $\kappa=10^{-6}$ and
    different step-sizes (from 0.3 to 1.1). Fig.\ref{mufig} demonstrates that
    exact reconstruction cannot be achieved at about $\mu=\{0.4,0.5,0.6,0.7,0.8\}$ with respective $M$ values,
    which are consistent with the values $\mu_{max}$ calculated by condition (\ref{mucond}).
    This result verifies our derivation in Theorem 1.

\section{Conclusion}
The adaptive filtering framework is introduced at the first time to
solve CS problem. Two typical adaptive filtering algorithms
$l_0$-LMS and $l_0$-EFWLMS, both imposing zero attraction method,
are introduced to solve CS problem, as well as to verify our
framework. In order to speed up the convergence of the two methods,
a novel algorithm $l_0$-ZAP, which adopts the zero attraction method
in the solution space, is further proposed. Thus the mean square
derivation of $l_0$-LMS in steady state has been deduced. The
performances of these methods have been studied experimentally.
Compared with those existing typical algorithms, they can
reconstruct signal with more nonzero coefficients under a certain
given number of measurements; while under a given sparsity, fewer
measurements are required by these algorithms. Moreover, they are
more robust against noise.

Up to now, there is no theoretical result for determining how to
choose the parameters of the proposed algorithms and how much the
number of measurements $M$ is in the context of RIP. These remain
open problems for our future work. In addition, our future work
includes the detailed discussion about the convergence performances
of $l_0$-EFWLMS and $l_0$-ZAP.

\appendix

\appendixtitleon
\begin{appendices}

\section{Proof of (\ref{noiserelation})}
\begin{proof}
The power of $\bf \hat{v}$ is
\begin{eqnarray} \label{vvhat}
  {\rm E} \left\{ {\bf \hat{v}}^{\rm T} {\bf \hat{v}} \right\} &=& {\rm E}\left\{ {\bf v}^{\rm T} {(\bf A^+)}^{\rm T} {\bf A^+v} \right\} \nonumber \\
   &=& {\rm E}\left\{ {\bf v}^{\rm T} \left[{\bf A}^{\rm T}{({\bf AA}^{\rm
   T})}^{-1}\right]^{\rm T} \left[{\bf A}^{\rm T}{({\bf AA}^{\rm
   T})}^{-1}\right] {\bf v}
   \right\}  \nonumber \\
   &=& {\rm E} \left\{ {\bf v} ({\bf AA}^{\rm T})^{-1} {\bf v}
   \right\} \nonumber \\
   &=& {\rm E} \left\{ {\bf v} {\rm E}\left\{({\bf AA}^{\rm T})^{-1}\right\} {\bf v}
   \right\}.
\end{eqnarray}
where the reason of the last equation of (\ref{vvhat}) holding is
that the noise $\bf v$ and measurement matrix $\bf A$ are
independent.

Suppose
\begin{equation}\label{Aaij}
    {\bf A}=\left( a_{ij} \right)_{1\leq i \leq M, 1\leq j \leq N}
\end{equation}
As mentioned in Section I, $a_{ij}$ is i.i.d. with
$\mathcal{N}(0,\frac{1}{M})$. Let
\begin{equation}\label{Bbij}
    {\bf B} = {\bf AA}^{\rm T}=(b_{ij})_{1\leq i \leq M, 1\leq j \leq
    M},
\end{equation}

Thus, for the diagonal components,
\begin{equation}\label{bii}
    b_{ii}=\sum_{k=1}^{N} {a_{ik}^2}, 1\leq i \leq M.
\end{equation}

Since $N$ is very large in CS, according to the central limit
theorem \cite{clt}, the following equation holds approximately,
\begin{equation}\label{bii1}
    b_{ii}\sim \mathcal{N}\left( {\rm E}{b_{ii}}, {\rm D}{b_{ii}}
    \right)=\mathcal{N}\left( \frac{N}{M}, \frac{2N}{M^2}
    \right)
\end{equation}
where D$\{\cdot\}$ denotes the variance. Similarly, for the
non-diagonal components,
\begin{equation}\label{bij}
    b_{ij}\sim \mathcal{N}\left( {\rm E}{b_{ij}}, {\rm D}{b_{ij}}
    \right)=\mathcal{N}\left( 0, \frac{N}{M^2}
    \right),i\neq j.
\end{equation}

Because ${N}/{M}\gg {2N}/{M^2}$, we have
\begin{equation}\label{Eaa}
    {\bf AA}^{\rm T}\approx \frac{N}{M}{\bf I}.
\end{equation}

Thus
\begin{equation}\label{Eaai}
    ({\bf AA}^{\rm T})^{-1}\approx \frac{M}{N}{\bf I}.
\end{equation}

Therefore equation  (\ref{vvhat}) can be simplified as
\begin{equation}\label{vvhat1}
   {\rm E} \left\{ {\bf \hat{v}}^{\rm T} {\bf \hat{v}} \right\}
   \approx \frac{M}{N}{\rm E} \left\{ {\bf v}^{\rm T} {\bf v} \right\}
\end{equation}

\end{proof}

\section{Proof of Theorem 1}
\begin{proof}
For simplicity, we use ${\bf w}(n)$, ${\bf x}(n)$, and $d(n)$
instead of ${\bf s}(k)$, ${\bf a}_k$ and $y_k$, respectively.
Suppose that $\bf w_o$ is the Wiener solution, thus
\begin{equation}\label{dn}
    d(n)={\bf x}^{\rm T}(n){\bf w_o}+v(n),
\end{equation}
where $v(n)$ is the measurement noise with zero mean. Define the
misalignment vector as
\begin{equation}\label{hn}
    {\bf h}(n) = {\bf w}(n)-{\bf w_o}.
\end{equation}

Thus we have
\begin{equation}\label{en}
    e(n)=v(n)-{\bf x}^{\rm T}(n){\bf h}(n).
\end{equation}

Equation (\ref{l0LMSrecu1}) is equivalent to
\begin{equation}\label{e1}
    {\bf h}(n+1) = \left[{\bf I}-\mu{\bf x}(n){\bf x}^{\rm T}(n)\right]{\bf
    h}(n)+\kappa{\bf g}({\bf w}(n)) + \mu v(n){\bf x}(n)
\end{equation}
Postmultiplying both sides of (\ref{e1}) with their respective
transposes,
\begin{eqnarray} \label{hh}
  {\bf h}(n+1){\bf h}^{\rm T}(n+1) \!\!\!&=&\!\!\! \left[{\bf I}-\mu{\bf x}(n){\bf x}^{\rm T}(n)\right]{\bf h}(n){\bf h}^{\rm T}(n)\left[{\bf I}-\mu{\bf x}(n){\bf x}^{\rm T}(n)\right]^{\rm T} \nonumber \\
   & &+\left[{\bf I}-\mu{\bf x}(n){\bf x}^{\rm T}(n)\right]{\bf h}(n)\kappa {\bf g}({\bf
   w}(n)) \nonumber \\
  & &+ \mu v(n)\left[{\bf I}-\mu{\bf x}(n){\bf x}^{\rm T}(n)\right]{\bf h}(n){\bf x}^{\rm T}(n) \nonumber \\
   & &+\kappa {\bf g}({\bf w}(n)){\bf h}^{\rm T}(n)\left[{\bf I}-\mu{\bf x}(n){\bf x}^{\rm T}(n)\right]^{\rm T} \nonumber \\
   & &+ \kappa^2 {\bf g}({\bf w}(n)){\bf g}^{\rm T}({\bf w}(n)) \nonumber \\
   & &+\kappa \mu v(n){\bf g}({\bf w}(n)){\bf x}^{\rm T}(n) \nonumber \\
   & &+ \mu v(n){\bf x}(n){\bf h}^{\rm T}(n)\left[{\bf I}-\mu{\bf x}(n){\bf x}^{\rm T}(n)\right]^{\rm T} \nonumber \\
   & &+\mu \kappa v(n){\bf x}(n){\bf g}^{\rm T}({\bf w}(n)) \nonumber \\
  & &+ \mu v^2(n){\bf x}(n){\bf x}^{\rm T}(n).
\end{eqnarray}

Let
\begin{equation}\label{Kn}
    {\bf K}(n)={\rm E}\left\{{\bf h}(n){\bf h}^{\rm T}(n)\right\}
\end{equation}
denote a second moment matrix of the coefficient misalignment
vector. Taking expectations on both sides of (\ref{hh}) and using
the Independence Assumption \cite{Haykin}, there is
\begin{align}\label{e2}
    {\bf K}(n+1) =&{\bf K}(n)-\mu \left({\bf RK}(n) + {\bf K}(n){\bf R}\right)+ 2\mu^2{\bf RK}(n){\bf R}
     \nonumber \\
    & + \mu^2{\bf R}\text{tr}\left({\bf RK}(n)\right)+2({\bf I}-\mu{\bf
    R})\kappa {\rm E}\left\{ {\bf h}(n){\bf g}^{\rm T}({\bf w}(n)) \right\} \nonumber \\
    &  +\kappa^2{\rm E}\{ {\bf g}\left({\bf w}(n)\right){\bf g}^{\rm T}\left({\bf w}(n)\right) \}+\mu^2 P_0{\bf
    R},
\end{align}
where
\begin{equation}\label{Rxx}
    {\bf R}={\rm E}\left\{ {\bf x}(n){\bf x}^{\rm T}(n)\right\}
\end{equation}
is the input correlation matrix,
\begin{equation}\label{p0}
    P_0={\rm E}\left\{v^2(n)\right\}
\end{equation}
is the minimum mean-squared estimation error and tr$\{\cdot\}$
denotes the trace.

As mentioned in Section I, $\bf A$ is i.i.d. Gaussian with mean zero
and variance $1/M$. Then
\begin{equation}\label{e3}
    {\bf R}= \frac{1}{M}{\bf I}.
\end{equation}

Therefore equation (\ref{e2}) can be simplified as
\begin{eqnarray}\label{e4}
    {\bf K}(n+1) \!\!\!&=&\!\!\!(1-\frac{2\mu}{M} +\frac{2\mu^2}{M^2}){\bf K}(n)
     + \frac{\mu^2}{M^2} \text{tr}\{{\bf K}(n) \}{\bf I}+2(1-\frac{\mu}{M})\kappa {\rm E}\left\{ {\bf h}(n){\bf g}^{\rm T}({\bf w}(n))\right\}\nonumber \\
    & & +\kappa^2{\rm E}\left\{ {\bf g}({\bf w}(n)){\bf g}^{\rm T}({\bf w}(n)) \right\}+\frac{\mu^2}{M} P_0{\bf
    I}.
\end{eqnarray}

Let
\begin{equation}\label{e5}
    D(n)={\rm E}\left\{\|{\bf w}(n)-{\bf w_o}\|_2^2\right\}=\text{tr}\left\{ {\bf K}(n)\right\}.
\end{equation}
Take the trace on both side of (\ref{e4}),
\begin{equation}\label{e6}
    D(n+1)=\left[1-\frac{2 \mu}{M}+\frac{(N+2)\mu^2}{M^2}\right]D(n) 
    +2(1-\frac{\mu}
    {M})\kappa \alpha (n)+\kappa^2 \beta(n) + \frac{N\mu^2}{M} P_0,
\end{equation}
where
\begin{equation}\label{e7}
    \alpha(n) = {\rm E}\left\{{\bf h}^{\rm T}({\bf w}(n)){\bf g}({\bf w}(n))\right\};
\end{equation}
\begin{equation}\label{e8}
    \beta(n) = {\rm E}\left\{{\bf g}^{\rm T}({\bf w}(n)){\bf g}({\bf w}(n))\right\}.
\end{equation}

Note that both $\alpha(n)$ and $\beta(n)$ are bounded,
\begin{align} 
  |\alpha(n)| &= \left|{\rm E}\left\{({\bf w}(n)-{\bf w_o}){\bf g}({\bf w}(n))\right\}\right| \nonumber \\
   &\leq  {\rm E}\left|\left\{({\bf w}(n)-{\bf w_o}){\bf g}({\bf w}(n))\right\}\right| \nonumber \\
   &\leq \sum_{i=0}^{N-1}{\rm E}\left|\left\{(w_i(n)- w_{oi})g(w_i(n))\right\}\right| \nonumber\\
   &= \sum_{|w_i(n)|<\frac{1}{\alpha}}{\rm E}\left|(w_i(n)- w_{oi})g(w_i(n))\right| \nonumber\\
  &\leq \sum_{|w_i(n)|<\frac{1}{\alpha}}{\rm E}\left\{|(w_i(n)- w_{oi})||g(w_i(n))|\right\} \nonumber\\
  &\leq \sum_{|w_i(n)|<\frac{1}{\alpha}}\alpha{\rm E}\left\{|(w_i(n)-
  w_{oi})|\right\} \quad(\because |g(w_i(n))|<\alpha)  \nonumber \\
  &\leq \sum_{|w_i(n)|<\frac{1}{\alpha}}\alpha\left\{{\rm E}|w_i(n)|+\|{\bf
  w_o}\|_1\right\} \nonumber \\
   &\leq N+\alpha\|{\bf w_o}\|_1;\label{alphabound}\\
  |\beta(n)| &= |{\rm E}\left\{{\bf g}^{\rm T}({\bf w}(n)){\bf g}({\bf w}(n))\right\}| \nonumber \\
   &\leq {\rm E}\left\{|{\bf g}^{\rm T}({\bf w}(n)){\bf g}({\bf w}(n))|\right\} \nonumber \\
  &\leq \sum_{i=0}^{N-1}{\rm E}\left\{|g(w_i(n))|^2\right\} \nonumber \\
   &\leq N\alpha^2.\label{betabound}
\end{align}

Therefore the following equation should be satisfied to guarantee
convergence of (\ref{e6}),
\begin{equation}\label{convcondition}
    |1-\frac{2 \mu}{M}+\frac{(N+2)\mu^2}{M^2}|<1.
\end{equation}
We have
\begin{equation}\label{mucond1}
    0<\mu<\frac{2M}{N+2}.
\end{equation}

The final mean square derivation in steady state is
\begin{equation}\label{finalmsd}
    D(\infty)=C\left[2\kappa (1-\frac{\mu}{M})\alpha (\infty)+\kappa^2 \beta(\infty) + \frac{N\mu^2}{M}
    P_0\right].
\end{equation}
where
\begin{align}
    C&=\frac{M^2}{2 \mu M-(N+2)\mu^2};\label{e9}\\
    \alpha(\infty) &= {\rm E}\left\{{\bf h}^{\rm T}({\bf w}(\infty)){\bf g}({\bf
    w}(\infty))\right\};\label{alphainfty}\\
    \beta(\infty) &= {\rm E}\left\{{\bf g}^{\rm T}({\bf w}(\infty)){\bf g}({\bf
    w}(\infty))\right\}.\label{betainfty}
\end{align}
\end{proof}

\end{appendices}

\section*{Acknowledgment}
The authors are very grateful to Mr. Detao Mao at the University of
British Columbia for his part in improving the English expression of
this paper. The authors also would like to express their cordial
thanks to the anonymous reviewers for their valuable comments on
this paper.

\end{document}